\documentclass[twocolumn,preprintnumbers,amsmath,amssymb,pra]{revtex4}

\usepackage[colorlinks,linkcolor=blue,anchorcolor=blue,citecolor=blue]{hyperref}
\usepackage{txfonts}
\usepackage{graphicx,hyperref}% Include figure files\underline{}
\usepackage{subfigure}
\usepackage{dcolumn}% Align table columns on decimal point
\usepackage{bm}% bold math
\usepackage{braket}
\usepackage{upgreek}
\usepackage{extarrows}
\usepackage{setspace}
\usepackage{float}
\usepackage{makecell}
\usepackage{rotating} 
\begin{document}
	\title{Spin-orbit coupling mediated photon-like resonance for a single atom trapped in a symmetric double well }
	\author{Changwei Fan$^{1}$}
	\author{Xiaoxiao Hu$^{1}$}
	\author{Xin Yan$^{1}$}
	\author{Hongzheng Wu$^{1}$}
	\author{Zhiqiang Li$^{1}$}
	\author{Jinpeng Xiao$^{2}$}
	\author{Yajiang Chen$^{1}$}
	\author{Xiaobing Luo$^{1,2}$}
	\altaffiliation{Corresponding author: xiaobingluo2013@aliyun.com}
	\affiliation{$^{1}$Department of Physics, Zhejiang Sci-Tech University, Hangzhou, 310018, China}
	\affiliation{$^{2}$School of Mathematics and Physics, Jinggangshan University, Ji’an 343009, China}
	\date{\today}% It is always \today, today%  but any date may be explicitly specified
	\begin{abstract}
		We employ a method involving coherent periodic modulation of Raman laser intensity to induce resonance transitions between energy levels of a spin-orbit coupled atom in a symmetric double-well trap. By integrating photon-assisted tunneling (PAT) technique with spin-orbit coupling (SOC), we achieve resonance transitions between the predefined energy levels of the atom, thereby enabling further precise control of the atom's dynamics. We observe that such photon-like resonance can induce a transition from a localized state to atomic Rabi oscillation between two wells, or effectively reduce tunneling as manifested by a quantum beating phenomenon. Moreover, such resonance transitions have the potential to induce spin flipping in a spin-orbit coupled atom. Additionally, the SOC-mediated transition from multiphoton resonance to fundamental resonance and the SOC-induced resonance suppression are also discovered.  In these cases, the analytical results of the effective coupling coefficients of the resonance transition derived from a four-level model can account for the entire dynamics, demonstrating surprisingly good agreement with the numerically exact results based on the realistic continuous model. 
		
	\end{abstract}
	\maketitle
	
	\section{Introduction}
	In recent years, periodic driving has emerged as a powerful tool in the field of cold atomic physics, allowing precise control and manipulation of atomic behavior\cite{1,2}. Periodic driving, also known as Floquet engineering, effectively facilitates the emulation and manipulation of  a wide range of quantum effects within cold atomic systems, such as exploring topological physics \cite{3,4,5,6,7}, controlling quantum tunneling \cite{8,9}, and manipulating spin dynamics \cite{10,11}. Floquet theory, as an effective tool for addressing problems involving time-periodic systems, is currently receiving a great deal of attention. This branch of quantum engineering is based on the principle that the time evolution of a periodically driven quantum system is governed by a time-independent effective Hamiltonian \cite{12,13,14}, apart from a micromotion described by a time-periodic unitary operator. By designing a suitable time-periodic driving protocol, researchers can engineer the properties of the effective Hamiltonian and achieve desired quantum phenomena  that are inaccessible in equilibrium systems. These include dynamic localization \cite{15,16}, the control of the bosonic superfluid-to-Mott-insulator transition \cite{17,18}, as well as the dynamic creation of kinetic frustration \cite{19,20}. Periodic driving has recently also been used to induce spin–orbit coupling \cite{21} and to realize a quantum ratchet \cite{22}.
	
	Tunneling control of ultracold atoms by periodic perturbations \cite{23,24,25,26} of potentials is currently established as an experimental method both at the single-particle level \cite{27} and at the level of Bose-Einstein condensates (BECs) \cite{18}. An interesting relevant effect is the analog of photon-assisted tunneling (PAT), which results from the energy exchange between the system and the oscillating field in integral multiples of photons \cite{8,28}. PAT is a resonance process that has been recognized as a powerful tool for the control of quantum tunneling \cite{8}. The investigation of PAT holds significant importance for understanding transport processes and for the development  of nanodevices \cite{28}. PAT stands out as a particularly fascinating and practical phenomenon, with a wide range of potential applications across diverse domains of physics \cite{29,30,31,32,33,34,35,36,37,38}. To date, it has been observed experimentally in semiconductor superlattices (\cite{29}), coupled quantum dots (\cite{31,32}), Josephson junctions (\cite{33}), and Bose-Einstein condensates in optical lattices \cite{36}. In non-Hermitian systems, PAT has also been used as a means to resonantly extend the domain of the $\mathcal{PT}$ symmetric phase \cite{39}. In particular, the phenomenon of PAT in BECs confined to double-well potentials has sparked interest  \cite{25,40,41}, and in this context, both integer-photon and fractional-photon resonances have been revealed and studied.
	
	The study of spin-orbit coupling (SOC) is an active area of research due to its ubiquitous occurrence in condensed matter phenomena such as topological insulators \cite{42,43}, the spin Hall effect \cite{44,45}, and spintronics \cite{46}. In recent years, experimental achievements have enabled the realization of artificial spin-orbit coupling in neutral bosonic systems \cite{47} and fermionic atomic gases \cite{48,49} through the coupling of two hyperfine states using Raman lasers. As a result, the ground-state properties and dynamical behavior of spin-orbit coupled atomic systems have garnered significant attention \cite{50,51,52,53,54,Abdullaev2018}. The double-well potential serves as a fundamental model for studying tunneling dynamics, and BECs in double-well potentials have been extensively investigated \cite{55,56,57}. Recent research has shown a growing interest in the dynamics of spin-orbit coupled BECs in double-well potentials \cite{58,59,60,61,62,63,64,65}. While many studies have primarily focused on the simplified two-site models, there are very few investigations into continuous models. Recently, there has been a significant contribution to understanding the suppression of tunneling in a spin-orbit coupled atom in a driven double-well system, achieved through the use of a realistic continuous model \cite{9}. Nevertheless, there has been limited research on the impact of the PAT effect on the dynamics of a spin-orbit coupled atom in double wells. On the other hand, the influence of periodically modulated Raman coupling has been studied both experimentally and theoretically in the context of spin-orbit coupled BECs \cite{66,67}. Motivated by these observations, this study aims to investigate the excitation of the PAT effect through the periodic modulation of Raman coupling, and to explore how the integration of PAT with SOC can be employed to achieve resonance transitions between specified energy levels, thus enabling precise control of the dynamics of the spin-orbit coupled atom trapped in a double-well potential. Additionally, we also delve deeper into studying the influence of SOC on the PAT mechanism using both the continuous model and the simplified four-state model.
	
	This paper is organized as follows. In Sec.~\ref{II}, we formulate the model and present the lowest four eigenstates and eigenvalues of the unperturbed system. We numerically demonstrate how the $\mathcal{PT}$ symmetry of the unperturbed eigenstates and the energy spacing between the energy levels change with the strength of the SOC. In Sec.~\ref{III}, we utilize these four eigenstates as a basis set, resulting in a simplified four-state model. Employing the degenerate perturbation theory in the extended Hilbert space, we derive an effective two-level Schr\"{o}dinger equation that precisely defines the effective coupling coefficients between the resonance states. All the resonance dynamics obtained from the realistic continuous model can be perfectly explained by the two-level Schr\"{o}dinger equation. In Sec.~\ref{IV}, we report on the dependence of the effective coupling strength of the resonance transition on the SOC strength. We observe that at certain values of SOC, the resonance transition can be completely suppressed, even when the fundamental resonance condition is met. Furthermore, we also discover that the transition from multiphoton resonance to fundamental (one-photon) resonance can be achieved by adjusting the SOC strength at a fixed low modulation frequency, thereby enhancing the strength of the resonance transition. In Sec.~\ref{V},  we summarize our results.

	\section{ THE CONTINUOUS MODEL}\label{II}
	
	We consider a single ultracold atom (or a noninteracting Bose-Einstein condensate) with two hyperfine pseudospin states $\ket{\uparrow}$ and $\ket{\downarrow} $, described by the spinor $\Psi = (\Psi_1, \Psi_2)^\mathsf{T}$, which is trapped in a one-dimensional symmetric double-well potential $V(x)$. The double-well potential, which is a combination of two identical even potentials, can be expressed as $V(x) = V_0(x + \frac{d}{2}) + V_0(x - \frac{d}{2})$ \cite{9}, where $V_0(x)=-U\text{exp}(-x^6/a^6)$ is an even function that is symmetric with respect to the transformation $x\rightarrow-x$. By adjusting the parameters $U$, $d$ and $a$, we can precisely control the depth and separation of the potential wells, as well as the tunneling rate between them. This enables a comprehensive study of quantum tunneling, energy level splitting, and other phenomena in the double-well potential system. The Raman laser field is used to generate a momentum-sensitive coupling between two internal atomic states, thereby realizing synthetic SOC in atomic systems \cite{47,48,49}. We assume that the Raman coupling is periodically varying in time, with $\Omega(t) = \Omega_0 + \Omega_1 \cos(\omega t) $, which can be easily realized in experiments by varying the Raman laser intensities \cite{68}. Here, $\omega$ is the modulation frequency, and $\Omega_1$ is the amplitude of modulation. Denoting the strength of the spin-orbit coupling by $\gamma$ and using the dimensionless unit $\hbar = M = 1$, the Hamiltonian of our continuous model takes the form $\hat{H} = \hat{H}_0 + \Omega_1 \cos(\omega t){\sigma}_x$, where
	\begin{equation}\label{eq1}
		\hat{H}_0 = \frac{\hat{p}^2}{2} - \gamma {\sigma}_z \hat{p} + \Omega_0{\sigma}_x + V({x}),
	\end{equation}
	is the unperturbed Hamiltonian of a spin–orbit-coupled atom in a double-well potential. Here, $\hat{p} = -i\frac{\partial}{\partial x}$ is the linear momentum operator, and ${\sigma}_{x,y,z}$ are the Pauli matrices.
	
	We assume that, for each well potential $V_0$, the two-component spin-orbit-coupled atom has two discrete eigenstates $\ket{i}$ ($i = 1,2$), which correspond to the two lowest energy levels of the single well. 
	\begin{figure}[htp]
		\center
		\includegraphics[width=9cm]{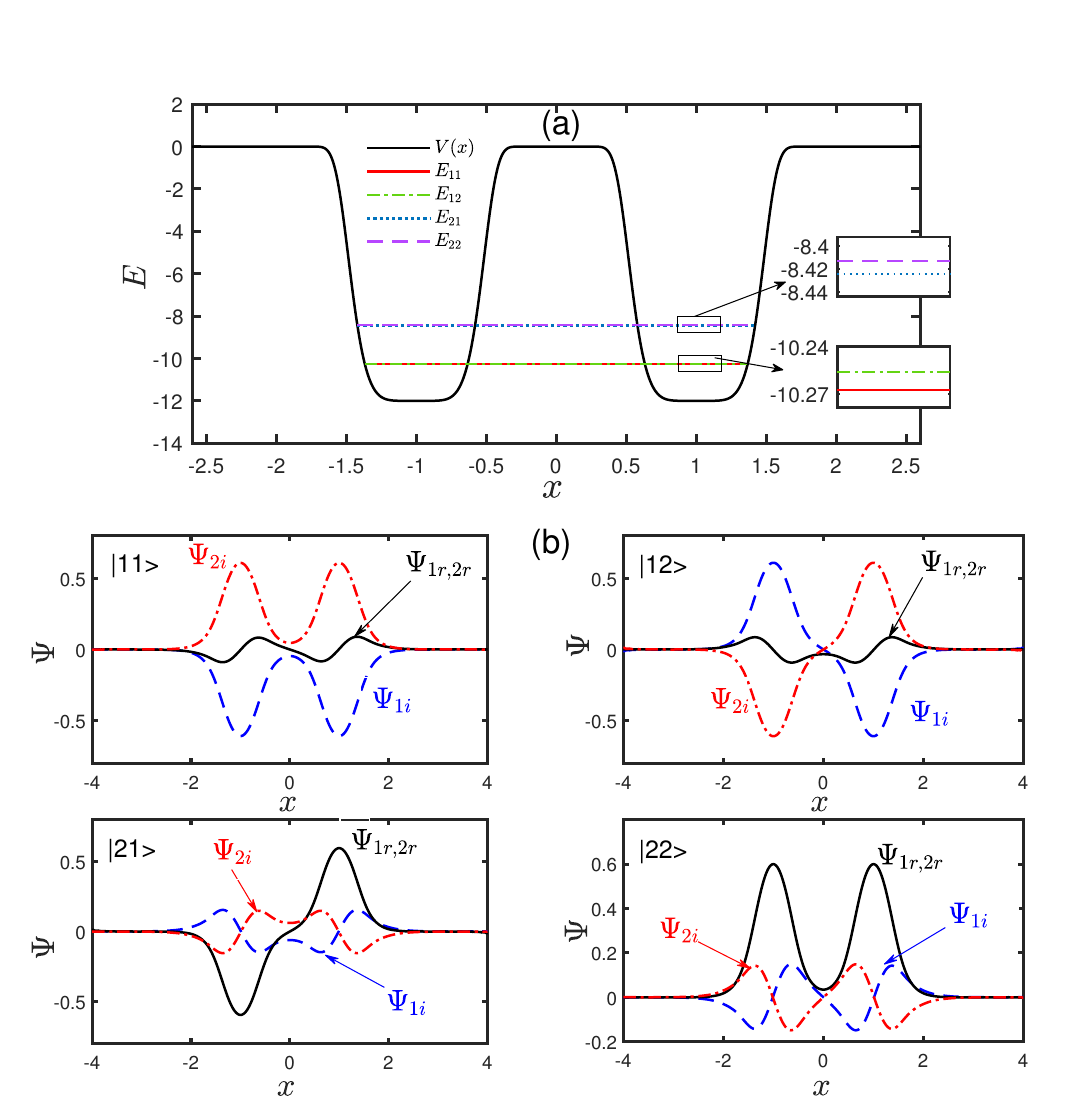}
		\caption{(a) The profile of the double-well potential, together with the four lowest discrete energy levels of $\hat{H}_0$. (b) The wavefunction profiles of the four eigenstates corresponding to the energy levels shown in (a). The subscripts $r$ and $i$ refer to the real and imaginary parts of the respective wavefunctions. The chosen parameters are $a=1/2$, $\Omega_0=1$, $\Omega_1=0.1$, $U=12$, $d=2$, and $\gamma=0.725$.}
		\label{fig1}
	\end{figure}
	The coupling with a neighboring well causes a splitting of these states, resulting in four stationary states $\ket{ij}$ ($i,j=1,2$) of the entire double-well potential, with $\hat{H}_0\ket{ij} = E_{ij}\ket{ij}$. In each state, the first index $i = 1,2$ corresponds to the lower and upper pair of levels, while the second index $j = 1,2$ corresponds to the lower and upper levels within each pair, as shown in Fig.~\ref{fig1}(a). According to this nomenclature, the energy levels are arranged in the following order: $E_{11} \leqslant E_{12} < E_{21} \leqslant E_{22}$, with all the energy levels being negative. The four eigenstates of $\hat{H}_0$ are shown in Fig.~\ref{fig1}(b). The corresponding modes are numerically obtained for the double-well potential  with $V_0(x) = -U \exp(-x^6/a^6)$, where the depth is $U=12$ and the width is $a=1/2$. The shape of the double-well potential and the corresponding energy levels of each eigenstate are shown in Fig.~\ref{fig1}(a). The distance between the two minima of the potential $V(x)$ is set to $d=2$, and $\Omega_0=1$.		
	The presence of SOC imposes significant constraints on the symmetries that the system can possess. The unperturbed Hamiltonian $\hat{H}_0$ respects three fundamental symmetries: $\hat{\alpha_1} = \mathcal{PT}$, $\hat{\alpha_2} = \sigma_x\mathcal{P}$, and $\hat{\alpha_3} = \sigma_x \mathcal{T}$, where $\mathcal{P}$ and $\mathcal{T}$ denote the parity and time reversal operators, respectively. These symmetry transformations, alongside the identity operator, have been verified to form a Klein four-group that is characterized by the relations $\hat{\alpha}_m\hat{\alpha}_n = \hat{\alpha}_n\hat{\alpha}_m = \hat{\alpha}_k$, for all indices. Hence, the eigenstates $\ket{ij}$ of $\hat{H}_0$ should be the eigenstates of the operators $\mathcal{PT}$, $\sigma_x\mathcal{P}$, and $\sigma_x\mathcal{T}$, adhering to following equation
	\begin{equation}\label{eq2}
		\mathcal{PT}\ket{ij}=\pm\ket{ij}, \sigma_x\mathcal{P}\ket{ij}=\pm\ket{ij}, \sigma_x\mathcal{T}\ket{ij}=\pm\ket{ij}.
	\end{equation}
	
	Through the exploitation of their symmetry properties, it can be shown that, for all eigenstates, the average $x$-component of spin is the only non-zero component. More specifically, this leads to the conclusion that $\langle ij | \sigma_x | ij \rangle \neq 0$, while both $\langle ij | \sigma_y | ij \rangle = 0$ and $\langle ij | \sigma_z | ij \rangle = 0$. 
	
\begin{figure}[htp]	
		\center
		\includegraphics[width=7cm]{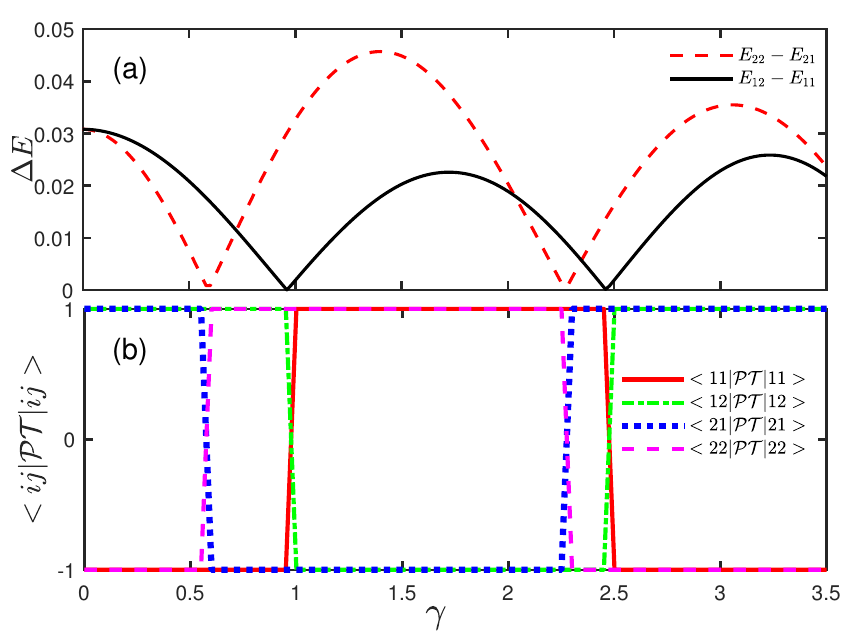}
		\caption{(a) The energy difference $\Delta E$ between eigenstates sharing the same index $i$  is plotted as a function of SOC strength $\gamma$. (b) The value of $\bra{ij}\mathcal{PT}\ket{ij}$ is shown as a function of the SOC strength $\gamma$. All other parameters are as previously defined in Fig.~\ref{fig1}.}
		\label{fig2}
	\end{figure}

	Next, we will explore how SOC affects and modifies the eigenenergy structure and $\mathcal{PT}$ symmetry. To do this, we numerically compute the energy gaps $E_{12} - E_{11}$ for the lower levels and $E_{22} - E_{21}$ for the upper levels as functions of the SOC strength. These calculations are visually represented in Fig.~\ref{fig2}(a). As the SOC strength increases, we observe that the energy levels of the lower and upper pairs approach degeneracy, with each pair reaching degeneracy at significantly different SOC strengths. Additionally, we investigate the interplay between the system's $\mathcal{PT}$ symmetry and the SOC by assessing the eigenvalues of the $\mathcal{PT}$ operator when acting on eigenstates of $\hat{H}_0$. These relationships are quantified by evaluating the values of  $\bra{ij}\mathcal{PT}\ket{ij}$, and the results are presented in Fig.~\ref{fig2}(b).
	It is observed that $\bra{ij}\mathcal{PT}\ket{ij}$ undergoes sharp transitions between the eigenvalues of $+1$ and $-1$, which are in one-to-one correspondence with the points at which the lower and upper pairs of energy levels collapse. The reason for this phenomenon is that, with the increase of SOC, the energy levels within a given pair cross, leading to the exchange of quantum states between the levels within that pair, which in turn causes a sudden transition in $\mathcal{PT}$ symmetry. 
	
	To set up the initial state, we introduce a new set of orthonormal basis states,
	\begin{equation}\label{eq3}
		\ket{i\pm}=\dfrac{1}{\sqrt{2}}(\ket{i2}\pm\ket{i1}).
	\end{equation}
	These states are not eigenstates of the Hamiltonian $\hat{H}_0$. Instead, the wavefunctions associated with $\ket{i\pm}$ are localized in either the left (-) or right (+) potential well of the double-well system. The symmetries imposed on the system indicate that the $x$-components of the mean spins for the localized modes, $S_{xi}=\bra{i\pm}\sigma_x\ket{i\pm}/2$, are directed oppositely for the lower ($i=1$) and upper ($i=2$) modes, yet are equal for modes localized in either the left or right well. Specifically, for $\gamma=0.725$ and $d=2$, we find that $S_{x1}=\bra{1\pm}\sigma_x\ket{1\pm}/2\approx-0.488$ and $S_{x2}=\bra{2\pm}\sigma_x\ket{2\pm}/2\approx0.47$.
	
	In this paper, unless otherwise specified, the initial state of the system is prepared as $\ket{1-}=\dfrac{1}{\sqrt{2}}(\ket{12}-\ket{11})$, which corresponds to the ground state wavefunctions of a single spin-orbit-coupled atom within an isolated left well potential.  In the absence of external perturbations, the state of the system evolves according to the equation $\hat{U}(t,0)\ket{1-}=\dfrac{1}{\sqrt{2}}(e^{-iE_{12}t}\ket{12}-e^{-iE_{11}t}\ket{11})$, where $\hat{U}(t,0)=e^{-i\hat{H}_0t}$ denotes the time evolution operator. In the unperturbed system, when $E_{11}$ is distinct from $E_{12}$, indicating a disparity in energy between the lower pair of levels, the system experiences a phenomenon known as tunneling. At specific times $t = (2n+1)\pi/(E_{12}-E_{11})$, where $n$ is an integer, the state $\ket{1-}$ (initially confined to the left well) transitions to the state $\ket{1+}$ (localized in the right well), signifying that the quantum particle has tunneled from one well to the other. This transition  happens periodically, and the tunneling period is determined as $T = 2\pi/(E_{12}-E_{11})$, which is the time it takes for the quantum particle to complete one full cycle of tunneling back and forth between the two wells. On the other hand, if $E_{11} = E_{12}$, indicating no energy difference between the two levels, the state $\ket{1-}$ remains localized in the left well and does not undergo tunneling. This is because the system is in a state of degeneracy, where both wells have the same energy, and the quantum particle does not switch between them. This implies that the energy splitting due to SOC offers new possibilities for controlling the quantum tunneling of particles in a double-well system.

	\section{ The impact of resonance transitions on dynamics  }\label{III}
	Application of an ac drive to the Raman coupling can significantly modify the tunneling dynamics of  a spin-orbit-coupled atom trapped in  a double-well potential. With the ac drive, $\Omega(t) = \Omega_0 + \Omega_1 \cos(\omega t) $,  the dynamics of the system are governed by  the time-dependent  Schr\"{o}dinger equation, 
	\begin{equation}\label{eq4}
		i\frac{\partial\boldsymbol{\Psi}(t)}{\partial t}=\hat{H}\boldsymbol{\Psi}(t)=[\hat{H}_0+\Omega_1 \cos(\omega t){\sigma}_x]\boldsymbol{\Psi}(t).
	\end{equation}
	
	To investigate the spin dynamics, we introduce the time-averaged probability,
	%we select the initial state of the system at $t=0$ to be $\ket{1-}$. Specifically, we have chosen $t = 40000$ in all numerical simulations. For rubidium atoms in the trap whose single-well width is 4 $\mu m$ (this corresponds to $a = 1/2$ in the super-Gaussian model used in our simulations), the dimensionless unit of time corresponds to approximately 1 $ms$ in the physical units. Thus averaging is performed over approximately 4 $s$. With sufficiently long time evolution,  we can use time-averaged quantities $\bar{P}$ and ${\bar{S}_n}$ to respectively describe the change in tunneling effects and spin direction.
	\begin{equation}\label{eq5}
		\begin{aligned}	
			&\bar{P}=\frac1{\Delta{t}}\int_0^{\Delta{t}}dtP(t),\\
			&P(t)=\int_{x_1}^{x_2}dx\boldsymbol{\Psi}^\dagger(x,t)\boldsymbol{\Psi}(x,t),	
		\end{aligned}
	\end{equation}
	and the time-averaged spin polarization,
	\begin{equation}\label{eq6}
		\begin{aligned}	
			&{\bar{S}_n}=\frac1{\Delta{t}}\int_0^{\Delta{t}}dt{S_n(t)},~~n = x, y, z,\\    
			&{S_n(t)}=\frac12\int_{-\infty}^{+\infty}dx\boldsymbol{\Psi}^\dagger(x,t){\sigma}_n\boldsymbol{\Psi}(x,t).
		\end{aligned}
	\end{equation}
	Here, $P(t)$ represents the probability of the atom being located in the left well [with $x_1=-\infty$ and $x_2=0$, hereafter referred to as $P_L(t)$] or the right well [with $x_1=0$ and $x_2=+\infty$, hereafter referred to as $P_R(t)$] at different times, and $\bar{P}$ signifies the time average of $P(t)$ over a sufficiently long time interval $\Delta t$. $S_n(t)$  with $n = x, y, z$ represents the spin polarization along the $x$, $y$, and $z$ axes, respectively. The time average of $S_n(t)$ is denoted by $\bar{S}_n$. We initialize the system at $t=0$ to $\ket{1-}$. In the dimensionless  Eq.~(\ref{eq4}), the energy is normalized in units of $\hbar \omega_0$, where $\omega_0 = \hbar {k^2}/{M}$ is the reference frequency. The length $x$ is scaled in units of $1/k$, and time $t$ is scaled in units of $1/\omega_0$. With these units, the dimensionless SOC strength is represented by $\gamma={k_R}/{k}$, where $k_R$ is the characteristic wave number that defines the SOC strength. Considering rubidium atoms confined in a double-well trap with a center-to-center distance of about $d = 2$  in our simulation, which is on the order of micrometers, the dimensionless time unit corresponds roughly to 1 millisecond in physical units. In the figures below, all frequencies are measured in units of the reference frequency $\omega_0 = 1$ kHz. To improve computational efficiency, the averaging time interval can be configured to cover a complete oscillation period corresponding to the longest period, which ensures that the data contains sufficient information during the averaging process. In all our simulations, the dimensionless time interval $\Delta t$ is less than 5000, indicating that the averaging time spans approximately several seconds. Additionally, the system parameters are fixed at $a=1/2, \Omega_0=1,\Omega_1=0.1,U=12$. %Consequently, the averaging time spans approximately 4 seconds. 
	
	%\begin{equation}\label{eq7}
	%\begin{aligned}		
	%&r_{c,n}(t)=\int_{-\infty}^{+\infty}dx\Psi^\dagger(x,t')(x\otimes\sigma_n)\Psi(x,t')\\
	%&{\bar{r}_{c,n}}=\frac1t\int_0^tdt'{r_{c,n}(t')},
	%\end{aligned}
	%\end{equation}
	%where $r_{c,n}(t)$  represents the spin center of mass at different times. ${\bar{r}_{c,n}}$ denotes the time average of  $r_{c,n}(t)$.
	
	By neglecting transitions to higher-energy levels, the system can be effectively modeled by using a finite-mode approximation, and the state \(\ket{\Psi(t)}\) at any given time \(t\) can be expressed as a linear combination of the basis states \(\ket{ij}\),
	\begin{equation}\label{eq7}
		\ket{\Psi(t)}=e^{-iE_0t}\sum_{i,j=1}^{2}c_{ij}(t)\ket{ij},
	\end{equation}
	where $\boldsymbol{\Psi}(x,t)=\langle{x}| \Psi(t)\rangle$ and $E_{0}=\frac{1}{4}\sum_{ij}E_{ij}$. The energy shift is chosen by subtracting \( E_0 \) to center the energy spectrum symmetrically around zero. Thus, the evolution of the column vector $\bm{c}=[c_{11},c_{12},c_{21},c_{22}]^\mathsf{T}$ is described by
	\begin{equation}\label{eq8}
		i\frac{d\mathbf{c}}{dt}=H\mathbf{c}=({H}_0+\Omega_1\cos(\omega t)\Gamma)\mathbf{c},
	\end{equation}
	with
	\begin{equation}\label{eq9}
		{H}_0\simeq\left[\begin{array}{cccc}
			E_{11}-E_{0} & 0 & 0 & 0\\
			0 & E_{12}-E_{0} & 0 & 0\\
			0 & 0 & E_{21}-E_{0}& 0\\
			0 & 0 & 0 & E_{22}-E_{0}\\
		\end{array}\right], 
	\end{equation}   
	and \(\Gamma\) is a \(4 \times 4\) matrix that characterizes the modulation. By assigning \(\ket{11} \equiv \ket{1}\), \(\ket{12} \equiv \ket{2}\), \(\ket{21} \equiv \ket{3}\), and \(\ket{22} \equiv \ket{4}\), the matrix elements of \(\Gamma\) are given by \(\Gamma_{m,n} = \bra{m}\sigma_x\ket{n}\) for \(m, n = 1, 2, 3, 4\).
	
	The probability \( P_{ij}(t) \) of finding the system to be in each eigenstate $\ket{ij}$ is expressed as
	\begin{equation}\label{eq10}
		\begin{aligned}		
			&P_{ij}(t)={\lvert{\langle{ij}| \Psi(t)\rangle}\rvert}^2={\lvert{c_{ij}(t)}\rvert}^2,\\
			%&\bar{P}_{ij}=\frac1{\Delta{t}}\int_0^{\Delta{t}}dtP_{ij}(t),\\
			%&P_{all}(t)=\sum_{i,j=1}^{2}P_{ij}(t)=1,
		\end{aligned}
	\end{equation}
	and the time average of \( P_{ij}(t) \) is given by
	\begin{equation}\label{11}
		\bar{P}_{ij}=\frac1{\Delta{t}}\int_0^{\Delta{t}}dtP_{ij}(t).
	\end{equation}
	Through direct numerical simulations of the Schr\"{o}dinger equation (\ref{eq4}), we have confirmed the validity of  $P_{all}=\sum_{i,j=1}^{2}P_{ij}(t)=1$, thereby showcasing the effectiveness of the  four-state approximation. Our subsequent theoretical analysis will be fundamentally grounded in the framework of the four-state model.
	
	Photon-assisted tunneling (PAT) is a resonant process known as a powerful tool for controlling quantum tunneling, arising from the exchange of \( m \) photons with the time-periodic (AC) field to bridge the energy gap between the lower-energy and higher-energy levels of the unperturbed system. As previously mentioned, the energy spectrum structure of the unperturbed system is influenced by the strength of SOC, which provides a novel avenue for us to control PAT using SOC. We are specifically interested in the manner in which SOC and PAT can be integrated to manipulate and control the dynamics of spin. In this paper, we focus our attention on the fundamental resonance, specifically when \(m=1\), exploring its implications and applications in depth. To gain a deeper understanding of the photon-like resonance, we utilized Floquet theory, with our analysis beginning in the extended Hilbert space. This space is spanned by the unperturbed Floquet states, denoted as \(\ket{m,ij} = e^{-im\omega t}\ket{ij}\). In the extended Hilbert space, the system's dynamics are governed by the Floquet Hamiltonian operator \(\hat{Q} = \hat{H}(t) - i\frac{\partial}{\partial t}\). The matrix representation of \(\hat{Q}\) in the basis of \(\ket{m,ij}\) is given by 
	
	\begin{equation}\label{12}
		Q=\left(\begin{array}{ccccc}
			\quad\ddots &\vdots &\vdots &\vdots &\begin{sideways}$\ddots$\end{sideways} \\
			\cdots & H_{\mathrm{0}}+\omega & H_{-1} &H_{-2} & \cdots\\
			\cdots	& H_{1} & H_{\mathrm{0}} & H_{-1} &\cdots \\
			\cdots	& H_2& H_{1} & H_{\mathrm{0}}-\omega &\cdots \\
			\begin{sideways}$\ddots$\end{sideways}&\vdots &\vdots &\vdots &\quad\ddots
		\end{array}\right),
	\end{equation}
	where the block matrices \( H_m \) are given by \( H_m =\frac{\omega}{2\pi} \int_0^{\frac{2\pi}{\omega}} e^{im\omega t} H dt \), with the nonzero matrices \( H_{-1} = H_{1} = \frac{\Omega_1}{2} \Gamma \) and \( H_0 \) defined by Eq.~(\ref{eq9}).
	
	We examine the first photon resonance ($m = 1$) with the resonance condition given by \( E_\beta - E_\alpha = \omega \), where the energy gap between the state \(\ket{\alpha} = \ket{1j}\) ($j=1,2$) in the lower pair of levels and the state \(\ket{\beta} = \ket{2j'}\) ($j'=1,2$) in the upper pair of levels is bridged by the energy of a single photon. For the sake of simplicity in notation, we use \(\ket{\alpha}\) and \(\ket{\beta}\) to represent the eigenstates of the lower and upper pairs of levels of the Hamiltonian \(\hat{H}_0\), respectively. In this case, the unperturbed Floquet states \(\ket{0, \alpha}\) and \(\ket{1, \beta}\) become degenerate, and we anticipate that the weak driving will mix these two resonant states. According to degenerate perturbation theory \cite{68,69,70} , the weakly driven system can be truncated to an effective two-level model operating in a reduced Hilbert space spanned by the Floquet states \(\{\ket{0,\alpha}, \ket{1,\beta}\}\). The corresponding effective \( Q_{\rm{eff}} \)-matrix is given by
	\begin{equation}\label{eq13}
		Q_{\rm{eff}}\simeq\left[\begin{array}{cc}
			\varepsilon_{\alpha} & \mathcal{V}\\
			\mathcal{V}^{\ast} & \varepsilon_{\beta}
		\end{array}\right],
	\end{equation}
	where \(\varepsilon_{\alpha} = E_{\alpha}\) and \(\varepsilon_{\beta} = E_{\beta} - \omega\) are the zeroth-order Floquet quasienergies with \(\varepsilon_{\alpha} = \varepsilon_{\beta}\), and \(\mathcal{V}\) is the first-order effective coupling coefficient, 
	\begin{equation}\label{eq14}
		\begin{aligned}\
			\mathcal{V}&=\bra{0,\alpha}{\Omega_1 \cos(\omega t){\sigma}_x}\ket{1,\beta}\\
			&=\frac{\bra{\alpha}\Omega_1\sigma_x\ket{\beta}}2\frac{\omega}{2\pi}\int_0^{\frac{2\pi}{\omega}}\left(e^{i\omega t}+e^{-i\omega t}\right)e^{-i\omega t}dt\\
			&=\frac{\Omega_1\bra{\alpha}\sigma_x\ket{\beta}}2.
		\end{aligned}
	\end{equation}
	Utilizing the symmetry properties of \( \hat{H}_0 \), we can demonstrate that the effective coupling coefficient \(\mathcal{V}\) is nonzero only when the states \(\ket{\alpha}\) and \(\ket{\beta}\) satisfy \(\bra{\alpha}\mathcal{PT}\ket{\alpha} = \bra{\beta}\mathcal{PT}\ket{\beta}\); otherwise, \(\mathcal{V} = 0\). This implies that a resonance transition can only occur if the two eigenstates of the unperturbed system share the same symmetry properties under the \(\mathcal{PT}\) operation. In other words, for resonance dynamics to take place, the two resonant eigenstates must possess the same eigenvalues under the \(\mathcal{PT}\) operation, meaning that both states are either symmetric (with an eigenvalue of $+1$) or antisymmetric (with an eigenvalue of $-1$) under \(\mathcal{PT}\) operation. If the states exhibit different \(\mathcal{PT}\) symmetries, they will not resonate. As numerically demonstrated in Fig.~\ref{fig2}(b), the \(\mathcal{PT}\) symmetry of the eigenstates is  influenced by  the SOC strength  \(\gamma\).
	
	At the fundamental resonance, the quantum dynamics of the system are described by the two-level Schr\"{o}dinger equation,
	\begin{equation}\label{eq15}
		i\frac\partial{\partial t}\bm{c}(t)=Q_{\rm{eff}}\bm{c}(t),
	\end{equation}
	where $\bm{c}(t)=[c_{0,\alpha}(t), c_{1,\beta}(t)]^\mathsf{T}$, with $c_{0,\alpha}(t)=\langle{0,\alpha}| \Psi(t)\rangle$ and $c_{1,\beta}(t)=\langle{1,\beta}| \Psi(t)\rangle$. When the system is initialized in the eigenstate \(\ket{\alpha}\) of the lower pair of energy levels, corresponding to the unperturbed Floquet state \(\ket{0, \alpha}\), the states \(\ket{0, \alpha}\) and \(\ket{1, \beta}\) will undergo Rabi oscillations, and their analytical solution is given by
	\begin{equation}\label{eq16}
		\begin{aligned}		
			&c_{0,\alpha}(t)=\cos(\lvert{ \mathcal{V}}\rvert t)e^{-i\varepsilon_{\alpha}t},\\
			&c_{1,\beta}(t)=i\sin(\lvert{ \mathcal{V}}\rvert t)e^{-i\varepsilon_{\beta}t}.
		\end{aligned}
	\end{equation}
	When returning to the conventional Hilbert space and using the relation \(\langle{1,\beta}| \Psi(t)\rangle = e^{i\omega t}\langle{0,\beta}| \Psi(t)\rangle\), we find that \(c_{\beta}(t) = c_{0,\beta}(t) = e^{-i\omega t}c_{1,\beta} = i\sin(\lvert{ \mathcal{V}}\rvert t)e^{-iE_{\beta}t}\).
	
\begin{figure}[htp]	
		\centering
		\includegraphics[width=6cm]{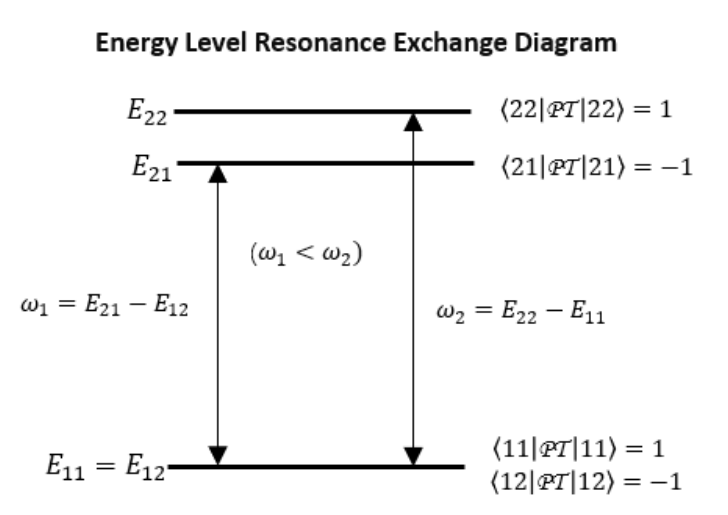}
		\caption{Schematic illustration of the resonance transition between unperturbed energy levels,  where $E_{11} =E_{12}$. The two resonance frequencies are designated as $\omega_1 = E_{21} - E_{12}$ and $\omega_2 = E_{22} - E_{11}$, with $\omega_1 < \omega_2$. It is noted that $\omega_1$ and $\omega_2$ are distinctly separated to prevent resonance overlap. The $\mathcal{PT}$ symmetry properties of the involved resonance states are specified, which are either symmetric (with eigenvalue $1$) or antisymmetric (with eigenvalue $-1$) under the $\mathcal{PT}$ operation. The resonance transition occurs exclusively between eigenstates with the same symmetry. This diagram is employed to schematically represent the resonance dynamics depicted in Fig.~\ref{fig4}. }
		\label{fig3}
	\end{figure}

	To study the resonance transition, we first consider a scenario in which the two unperturbed eigenstates \( |11\rangle \) and \( |12\rangle \), which combine to form the initial state \( |1-\rangle \), are degenerate in energy, i.e. \( E_{11} = E_{12} \). This degeneracy can be achieved by tuning the SOC as discussed earlier. The schematic representation of the resonant transfer between energy levels for this scenario is shown in Fig.~\ref{fig3}, where the two resonance frequencies are given by \(\omega_1 = E_{21} - E_{12}\) and \(\omega_2 = E_{22} - E_{11}\), respectively.  The resonance phenomena depicted in Fig.~\ref{fig3} are validated by directly solving the Schr\"{o}dinger equation (\ref{eq4}) using the split-step Fourier (SSF) method for the initial input state  \(\ket{1-}\) with the parameters $d=1.7$, $\gamma=1.112$ and $\Omega_1=0.1$.
	\begin{figure*}[tbp]
		\centering
		\includegraphics[scale=0.55]{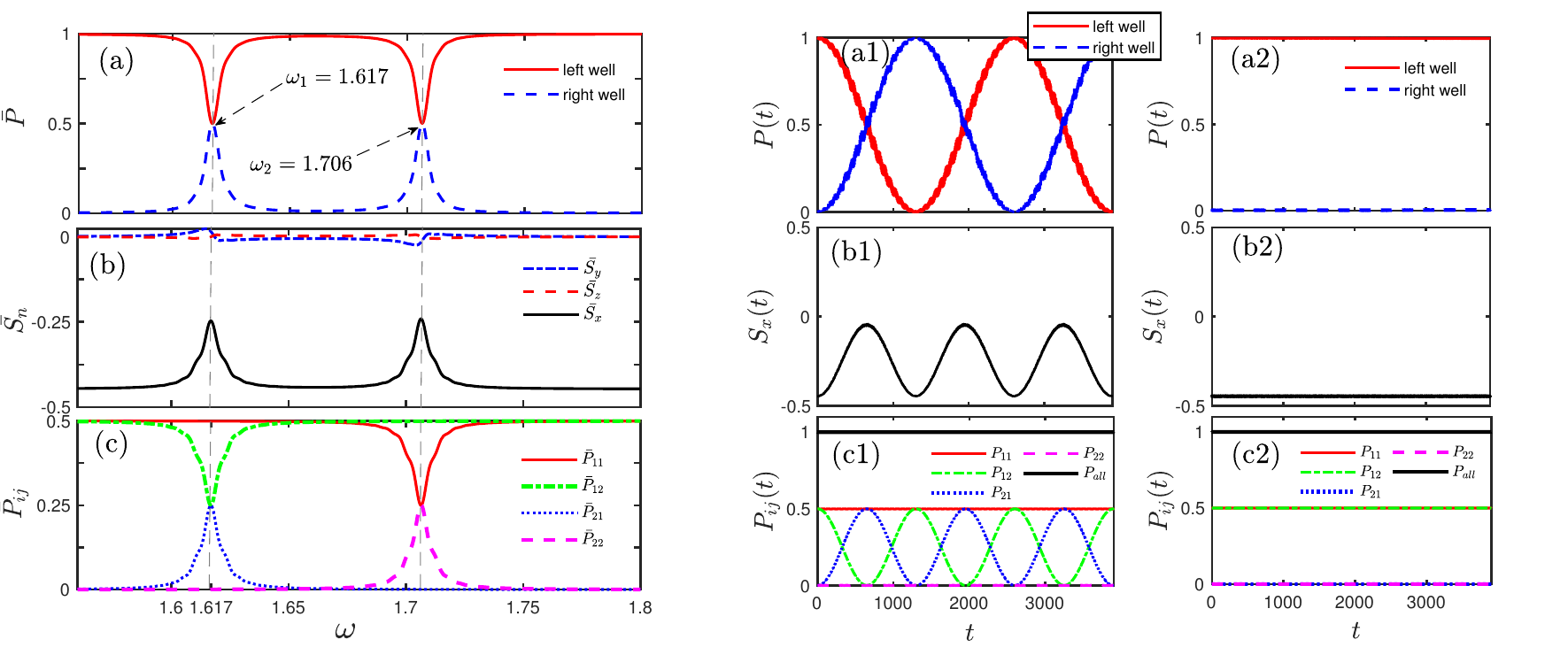}
		\caption{(a), (b), and (c) depict the time-averaged probability $\bar{P}$ of finding the atom in the left (or right) well, the time-averaged spin polarization $\bar{S}_n$ in the three orthogonal ($n=x, y, z$) directions, and the time-averaged probability $\bar{P}_{ij}$ of the system being in each eigenstate $\ket{ij}$, respectively, as functions of the modulation frequency $\omega$. (a1), (b1), and (c1) show the plots  of $P(t)$, $S_x(t)$, and $P_{ij}$ as functions of time at the resonance frequency $\omega=\omega_1 = 1.617$. (a2), (b2), and (c2) show the plots of $P(t)$, $S_x(t)$, and $P_{ij}$ as functions of time at the non-resonant frequency $\omega = 1.655$. All the results presented are derived from numerical simulations of the Schr\"{o}dinger equation (\ref{eq4}). The other parameters are $d = 1.7$, $\gamma = 1.112$, and $\Omega_1 = 0.1$.}
		%\caption{(a), (b), and (c) depict the variations of $\bar{P}$, ${\bar{S}_n}$, and ${\bar{P}_{ij}}$ with respect to the driving frequency $\omega$ . (a1), (b1), and (c1) display the variations of $P(t)$, ${S_x(t)}$, and $P_{ij}$ with respect to time $t$ at the resonance frequency ($\omega_1=1.617$). (a2), (b2), and (c2) display the variations of $P(t)$, $S_x(t)$, and $P_{ij}$ with respect to time $t$ at the non-resonance frequency ($\omega=1.655$). The parameters are set to $d=1.7$, $\gamma=1.112$, $\Omega_1=0.1$.}
		\label{fig4}
	\end{figure*}
	In Figs.~\ref{fig4}(a), (b), and (c), we present the dependence of $\bar{P}$, ${\bar{S}_n}$, and ${\bar{P}_{ij}}$ on the modulation frequency $\omega$, which reveals two pronounced resonance peaks at $\omega_1 = 1.617$ and $\omega_2 = 1.706$. We have numerically confirmed that these two resonance peaks correspond to the one-photon process at frequencies of \(\omega_1=E_{21}-E_{12}\) and \(\omega_2=E_{22}-E_{11}\),  where \(E_{11}=E_{12}\). We also compared the time-dependent behavior of $P(t)$, ${S_n(t)}$, and $P_{ij}(t)$ at the resonance frequency $\omega_1=1.617$ [see Figs.~\ref{fig4}(a1), (b1) and (c1)] and at the non-resonant frequency $\omega=1.655$ [see Figs.~\ref{fig4}(a2), (b2) and (c2)]. At the non-resonant frequency $\omega=1.655$, the system exhibits frozen dynamics, with all physical observables maintaining their initial values. The occurrence of resonance causes a transition from the frozen dynamics to full tunneling between the two wells [see Fig.~\ref{fig4}(a1), where $P(t)$ can reach zero], and it results in Rabi oscillation of the $x$-direction spin polarization between approximately $-0.5$ and $0$ [see Fig.~\ref{fig4}(a2)]. By analyzing the time evolution of the probability \( P_{ij}(t) \) at the resonance frequency \(\omega_1=1.617\), it is evident that resonant transitions can only occur between the eigenstates \( |12\rangle \) and \( |21\rangle \) [see Fig.~\ref{fig4}(c1)]. Due to the well-separated resonance frequencies \(\omega_1\) and \(\omega_2\), neither resonance nor near-resonance occurs between \( |11\rangle \) and \( |22\rangle \) at \(\omega = \omega_1\). Moreover, the resonance transition between \( |11\rangle \) and \( |21\rangle \) is forbidden due to their contrasting \(\mathcal{PT}\) symmetries, with \(\bra{11}\mathcal{PT}\ket{11}=1\) and \(\bra{21}\mathcal{PT}\ket{21}=-1\). Thus, when the system is at the resonance frequency $\omega_1$,  $\ket{12}$ and $\ket{21}$ are coupled, and according to Eq.~(\ref{eq16}), the system evolves as follows  
	\begin{equation}\label{eq17}
		\begin{aligned}		
			\ket{\Psi(t)}=&\dfrac{1}{\sqrt{2}}(\cos(\lvert{ \mathcal{V}}\rvert t)e^{-iE_{12}t}\ket{12}-e^{-iE_{11}t}\ket{11}+\\&i\sin(\lvert{ \mathcal{V}}\rvert t)e^{-iE_{21}t}\ket{21}),
		\end{aligned}
	\end{equation}
	where $E_{11}= E_{12}$, and the state basis $\ket{11}$ evolves at its own frequency. From Eq.~(\ref{eq17}), we can obtain some relevant analytical results. At  \( t=\dfrac{(2n+1)\pi}{2\lvert{\mathcal{V}}\rvert} \), where \( n \) is an integer, we have \( S_{x}(t)=\dfrac{1}{2}\bra{\Psi(t)}\sigma_x\ket{\Psi(t)}\approx0 \), as can be observed in Fig.~\ref{fig4}(b1).
	\begin{figure}[htp]	%共振示意图2
		\centering
		\includegraphics[width=6cm]{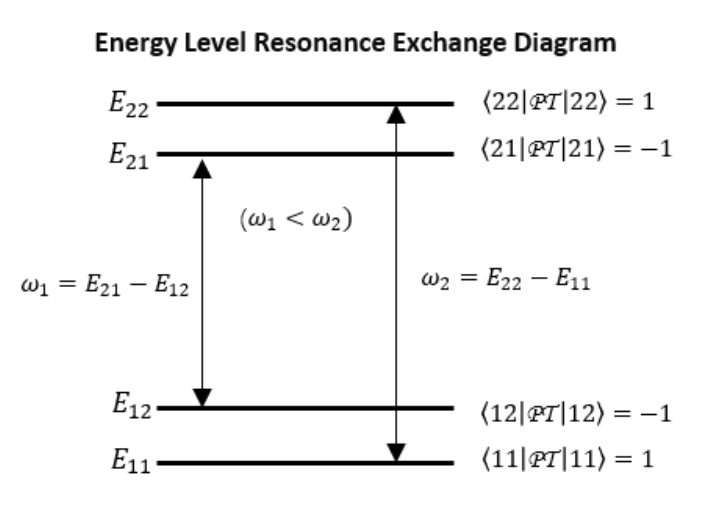}
		\caption{Schematic representation of the resonance transition between unperturbed energy levels, where $E_{11} \neq E_{12}$. The two resonance frequencies are defined as $\omega_1 = E_{21} - E_{12}$ and $\omega_2 = E_{22} - E_{11}$, with $\omega_1 < \omega_2$, which is consistent with Fig.~\ref{fig3}. The $\mathcal{PT}$ symmetry properties of the involved resonance states are also specified. This diagram is used to schematically illustrate the resonance dynamics shown in Fig.~\ref{fig6}.}
		\label{fig5}
	\end{figure}
	\begin{figure*}[hbtp]
		\centering
		\includegraphics[scale=0.55]{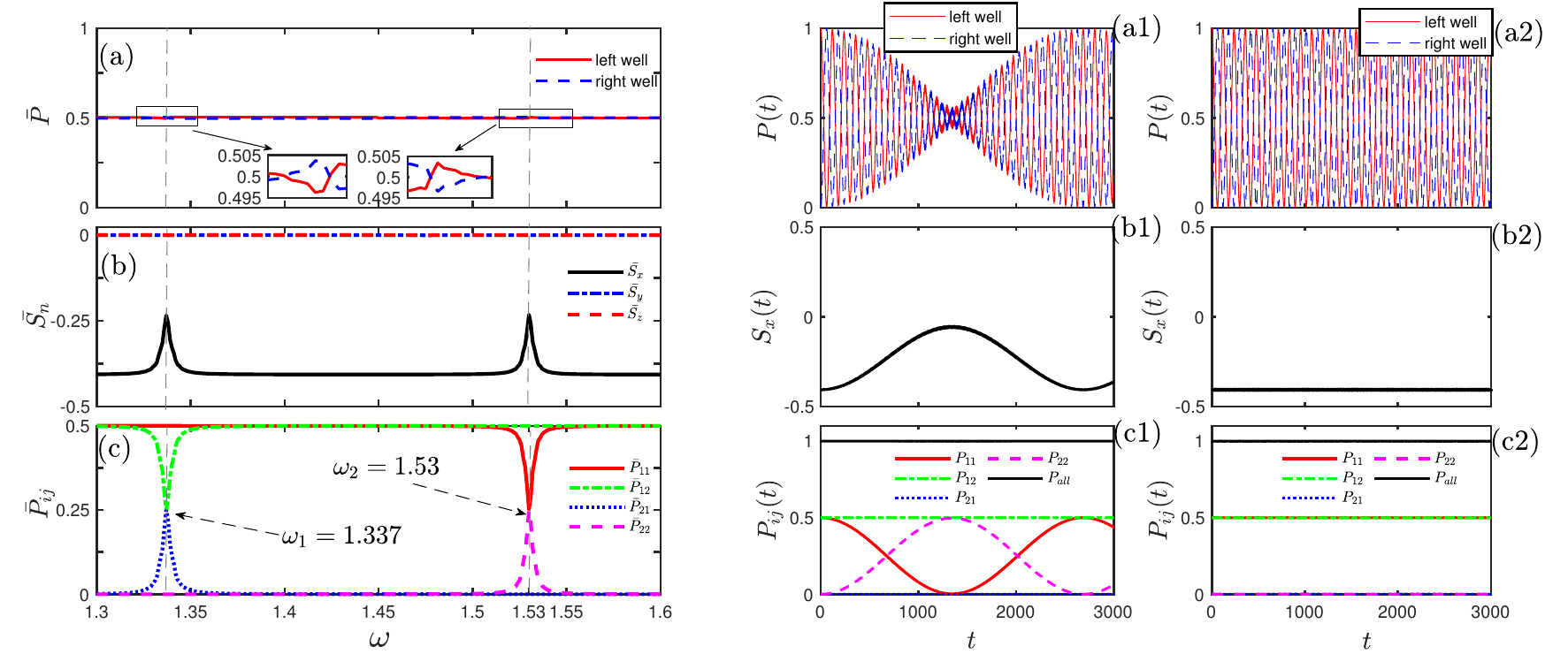}
		\caption{(a), (b), and (c) respectively show the time-averaged probability $\bar{P}$ of finding the atom in the left (or right) well, the time-averaged spin polarization $\bar{S}_n$ in the three orthogonal ($n = x, y, z$) directions, and the time-averaged probability $\bar{P}_{ij}$ of the system being in each eigenstate $\ket{ij}$, as functions of the modulation frequency $\omega$. (a1), (b1), and (c1) show the plots of $P(t)$, $S_x(t)$, and $P_{ij}(t)$ as functions of time at the resonance frequency $\omega=\omega_2 = 1.53$. (a2), (b2), and (c2) show the plots of $P(t)$, $S_x(t)$, and $P_{ij}(t)$ as functions of time at the non-resonant frequency $\omega = 1.45$. All the results presented are derived from numerical simulations of the Schr\"{o}dinger equation (\ref{eq4}). The other parameters used are $d = 1.7$, $\gamma = 1.5$, and $\Omega_1 = 0.1$.}
		\label{fig6}
	\end{figure*}
	This occurs because the states \( \ket{12} \) and \( \ket{21} \) have fully transitioned to each other at this moment, leading to the system's wavefunction being composed of two eigenstates, \( \ket{11} \) and \( \ket{21} \), with opposite spin polarization in the $x$-direction. At \( t=\dfrac{(2n+1)\pi}{\lvert{\mathcal{V}}\rvert} \), where \( n \) is an integer, the state of the atom  is \( \ket{\Psi(t)} \) =\( -\dfrac{1}{\sqrt{2}}(\ket{12} + \ket{11}) e^{-iE_{11}t}\)=\( -e^{-iE_{11}t}\ket{1+} \). This indicates that the atom has completely tunneled from the left well to the right well. Given the localized basis set \(\{\ket{i\pm}\}\), the probability of finding the atom in the left well at any given time can be expressed as:
	\begin{equation}\label{eq18}
		P_L(t)={\lvert{\langle1-| \Psi(t)\rangle}\rvert}^2+{\lvert{\braket{2- | \Psi(t)}}\rvert}^2.
	\end{equation}
	Inserting Eq.~(\ref{eq17}) into Eq.~(\ref{eq18}) yields
	\begin{equation}\label{eq19}
		P_L(t)=\dfrac{1}{2}+\dfrac{1}{2}\cos(\lvert{ \mathcal{V}}\rvert t),
	\end{equation}
	which perfectly accounts for the numerical results obtained from Eq.~(\ref{eq4}) as  illustrated  in Fig.~\ref{fig4}(a1).
	
\begin{figure}[htp]
		\center
		\includegraphics[width=8cm]{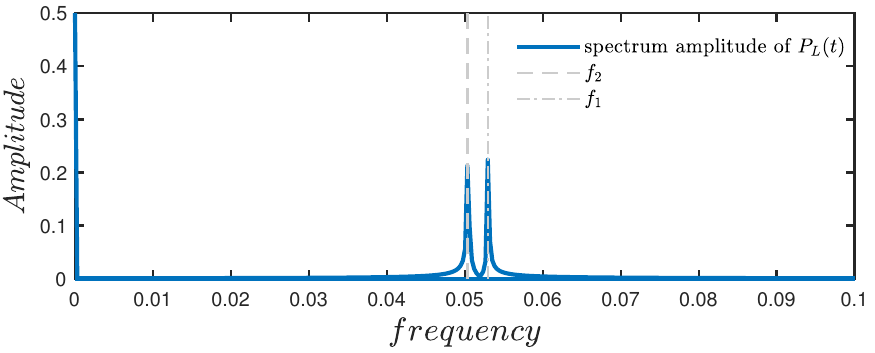}
		\caption{Frequency spectrum of $P_L(t)$ for the case in Fig.~\ref{fig6}(a1). The blue line depicts the result obtained from Fourier analysis of the numerical data for $P_L(t)$ in Fig.~\ref{fig6}(a1). The vertical (dashed and dash-dotted) lines in the figure correspond to two analytical frequencies obtained from Eq.~(\ref{eq21}).}
		\label{fig7}
	\end{figure}

	Second, we consider the case where the two unperturbed eigenstates \( |11\rangle \) and \( |12\rangle \) have different energies, \( E_{11} \neq E_{12} \). The corresponding schematic diagram of the energy level resonance transition is shown in Fig.~\ref{fig5}, where there is one resonance transition between \(|12\rangle\) and \(|21\rangle\) at a frequency of \(\omega_1 = E_{21} - E_{12}\), and another resonance transition between \(|11\rangle\) and \(|22\rangle\) at a frequency of \(\omega_2 = E_{22} - E_{11}\). In this case, we change the SOC parameter to \(\gamma=1.5\) and leave the other parameters the same as in Fig.~\ref{fig4}. The corresponding dynamical behavior obtained from Eq.~(\ref{eq4}) is shown in Fig.~\ref{fig6}. In Fig.~\ref{fig6}(a), (b) and (c) we show the dependence of $\bar{P}$, ${\bar{S}_n}$ and ${\bar{P}_{ij}}$ on the modulation frequency $\omega$. Two distinct resonance peaks at frequencies \(\omega_1 = 1.337\) and \(\omega_2 = 1.53\) are clearly observed in the time-averaged $x$-component spin polarization \(\bar{S}_x\) [see Fig.~\ref{fig6}(b)], accompanied by sharp transitions of \(\bar{P}_{ij}\) at the two resonance frequencies [see Fig.~\ref{fig6}(c)]. While at the frequencies \(\omega_1 = 1.337\) and \(\omega_2 = 1.53\) the time-averaged probabilities \(\bar{P}\) change only slightly. In particular, we have also examined the time-dependent behavior of \( P(t) \), \( S_n(t) \), and \( P_{ij}(t) \) for two typical modulation frequencies, one at the resonant frequency \(\omega_2 = 1. 53\) [see Figs.~\ref{fig6}(a1), (b1), and (c1)], and the other at the non-resonant frequency \(\omega = 1.45\) [see Figs.~\ref{fig6}(a2), (b2), and (c2)]. At the non-resonant frequency $\omega=1.45$, due to the non-degenerate energies of the eigenstates constituting the initial mode \(\ket{1-}\), the atom undergoes Rabi tunneling between the left and right wells, as shown in Fig.~\ref{fig6}(a2), which is similar to the behavior observed in the non-driven scenario. In this scenario, the state maintains an equal superposition of the two eigenstates \(\ket{11}\) and \(\ket{12}\) at the lower pairs of energy levels throughout the evolution process [see Fig.~\ref{fig6}(c2)], without transitioning to the upper pairs of energy levels. This lack of transition results in the spin polarization \(S_x\) remaining constant [see Fig.~\ref{fig6}(b2)]. In contrast, at the resonance frequency of \(\omega_2 = 1.53\), the time-dependent behavior of \(P(t)\) exhibits a pattern known as quantum beating  [see Fig.~\ref{fig6}(a1)]. In this pattern, the oscillation amplitude of \(P(t)\) decreases gradually until it reaches zero, and then it starts increasing again. We observe that the resonance frequency \(\omega_2\) corresponds to a one-photon resonance between \(\ket{11}\) and \(\ket{22}\), which is evident from the periodic exchange of \(P_{11}\) and \(P_{22}\) in Fig.~\ref{fig6}(c1). This resonance transition leads to the oscillation of \( S_x \) near -0.5 and 0, as shown in Fig.~\ref{fig6}(b1). Therefore, at the resonance frequency \(\omega_2\), the evolution of the system is described by
	\begin{equation}\label{eq20}
		\begin{aligned}		
			\ket{\Psi(t)}=&\dfrac{1}{\sqrt{2}}(e^{-iE_{12}t}\ket{12}-\cos(\lvert{ \mathcal{V}}\rvert t)e^{-iE_{11}t}\ket{11}+\\&i\sin(\lvert{ \mathcal{V}}\rvert t)e^{-iE_{22}t}\ket{22}),
		\end{aligned}
	\end{equation}
	where \(E_{11}\neq E_{12}\). At this resonant frequency, \(\ket{12}\) evolves at its own frequency, as there is no resonance transition between \(\ket{12}\) and other states. According to the results from Eq.~(\ref{eq20}), we can explain why the spin polarization \( S_x(t) \) is close to zero at the peak of the resonance transition. At times \( t=\dfrac{(2n+1)\pi}{2\lvert{\mathcal{V}}\rvert} \), where \( n \) is an integer, there is a full exchange between states \( \ket{11} \) and \( \ket{22} \). This exchange leaves the system mainly in the states \( \ket{12} \) and \( \ket{22} \), which have opposite spin polarizations in the $x$-direction. Since these states are antiparallel, their combination cancels out the total spin polarization along the $x$-axis, making \( S_{x}(t) \approx 0 \). By substituting Eq.~(\ref{eq20}) into Eq.~(\ref{eq18}), we obtain an explicit formula for the probability \( P_L(t) \) of finding an atom in the left well at any given time:
	\begin{equation}\label{eq21}
		P_L(t)=\dfrac{1}{2}+\dfrac{1}{4}\{(\cos[(\lvert{ \mathcal{V}}\rvert+\Delta{E}) t]+\cos[(\lvert{ \mathcal{V}}\rvert-\Delta{E}) t]\},
	\end{equation}
	with \( \Delta{E}=E_{12}-E_{11} \). The characteristics of \( P_L(t) \) can be described as a combination of cosinusoidal waves (oscillations) with different frequencies, denoted as \( f_1 = |\mathcal{V}| + \Delta{E} \) and \( f_2 = |\mathcal{V}| - \Delta{E} \). When the frequencies of the two vibrations are very close, a beat frequency phenomenon occurs, as described above. The existence  of these two slightly different frequencies leads to a substantial increase in the time it takes for the system to fully tunnel from the left well to the right well [as seen by comparing Figs.~\ref{fig6}(a1) and (a2)], which effectively weakens the tunneling effect to some extent. To validate the consistency between the analytical formula (\ref{eq21}) and the numerical data shown in Fig.~\ref{fig6}(a1), a frequency spectrum analysis was performed on the numerical results for \( P_L(t) \) as given by Eq.~(\ref{eq4}). We find that the spectrum amplitudes exhibit two peaks centered at the slightly shifted frequencies, which exactly correspond to the analytical frequencies \( f_1 \) and \( f_2 \), as shown in Fig.~\ref{fig7}. In general,  even in the presence of resonance transition, the long-time average of \( P_L(t) \) will converge to a value close to $0.5$, as shown in Fig.~\ref{fig6}(a). We can adjust the modulation amplitude $\Omega_1$ such that \(\lvert{ \mathcal{V}}\rvert=\Delta{E}\), thereby causing the probability \( P_L \) to be expressed as \( P_L(t)=\dfrac{3}{4}+\dfrac{1}{4}\cos[(\lvert{ \mathcal{V}}\rvert+\Delta{E}) t] \). This adjustment prevents the atom from tunneling completely to the right well. 
	
	\begin{figure}[htp]	
		\center
		\includegraphics[width=6cm]{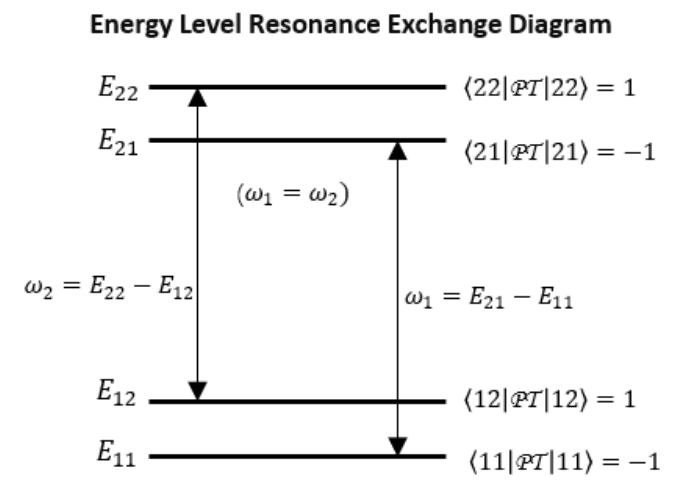}
		\caption{Schematic illustration of the resonance transition between unperturbed energy levels, where $E_{11} \neq E_{12}$, but with different resonance pathways from those in Fig.~\ref{fig5}. The two resonance frequencies are designated as $\omega_1 = E_{22} - E_{12}$ and $\omega_2 = E_{21} - E_{11}$, with $\omega_1 = \omega_2$. The $\mathcal{PT}$ symmetry properties of the involved resonance states are also specified. This diagram is used to schematically represent the resonance dynamics depicted in Fig.~\ref{fig9}. }
		\label{fig8}
	\end{figure}
	
	\begin{figure*}[tbp]
		\centering
		\includegraphics[scale=0.55]{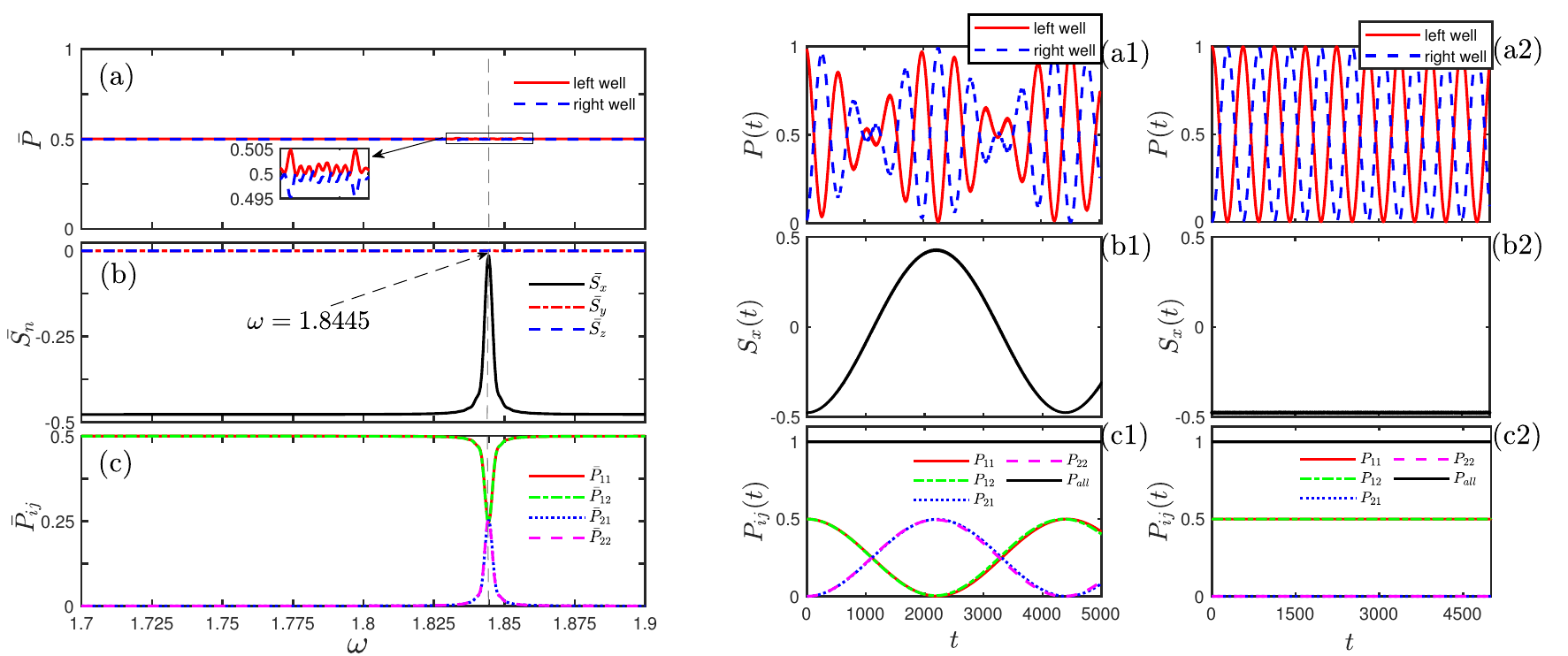}
		\caption{(a), (b), and (c) respectively show the time-averaged probability $\bar{P}$ of finding the atom in the left (or right) well, the time-averaged spin polarization $\bar{S}_n$ in the three orthogonal ($n = x, y, z$) directions, and the time-averaged probability $\bar{P}_{ij}$ of the system being in each eigenstate $\ket{ij}$, as functions of the modulation frequency $\omega$. (a1), (b1), and (c1) show the plots of $P(t)$, $S_x(t)$, and $P_{ij}(t)$ as functions of time at the resonance frequency $\omega=\omega_1=\omega_2 = 1.8445$. (a2), (b2), and (c2) show the plots of $P(t)$, $S_x(t)$, and $P_{ij}(t)$ as functions of time at the non-resonant frequency $\omega = 1.2$. All the results presented are derived from numerical simulations of the Schr\"{o}dinger equation (\ref{eq4}). The other parameters are $d=2$, $\gamma=0.725$, $\Omega_1=0.1$.}
		\label{fig9}
	\end{figure*}

	Thirdly, we continue to consider that the energies of the two eigenstates that constitute the initial state are distinct from each other, with \( E_{11} \neq E_{12} \). However, unlike in Fig.~\ref{fig5}, we consider a different scenario where \(\ket{11}\) and \(\ket{21}\) resonate at the frequency \(\omega_1\), while \(\ket{12}\) and \(\ket{22}\) resonate at the frequency \(\omega_2\), as schematically illustrated in Fig.~\ref{fig8}. In this scenario, we can fine-tune the system parameters to ensure that both \(\ket{11}\) and \(\ket{12}\) experience resonance at a single modulation frequency, such that \(\omega = \omega_1 = \omega_2\). The corresponding dynamics are depicted in Fig.~\ref{fig9} through numerical simulations of Eq.~(\ref{eq4}) with the specified parameters $d=2$, $\gamma=0.725$ and $\Omega_1=0.1$. The dependence of \(\bar{P}\), \({\bar{S}_n}\) and \({\bar{P}_{ij}}\) on the modulation frequency \(\omega\) is shown in Figs.~\ref{fig9}(a), (b) and (c), where we observe a single resonance peak in the time-averaged \(x\)-directional spin polarization \(\bar{S}_x\) (with the peak value of \(\bar{S}_x\) being close to zero). The single peak arises from the fact that the resonances between \(\ket{11}\) and \(\ket{21}\) and between \(\ket{12}\) and \(\ket{22}\) completely overlap, as shown in Fig.~\ref{fig9}(c) and Fig.~\ref{fig9}(c1). We also analyzed the time evolution of \(P(t)\), \({S_n(t)}\), and \(P_{ij}(t)\) at the resonance frequency \(\omega=1.8445\) [as depicted in Figs.~\ref{fig9}(a1), (b1) and (c1)] and compared it with the behavior at a non-resonance frequency of \(\omega=1.2\) [as shown in Figs.~\ref{fig9}(a2), (b2) and (c2)]. For the non-resonance frequency of \(\omega = 1.2\), the dynamics are similar to those depicted in Figs.~\ref{fig6}(a2), (b2), and (c2). At the resonance frequency \(\omega = 1.8445\), the probability function \(P(t)\) exhibits a quantum beating  phenomenon as shown in Fig.~\ref{fig9}(a1). Specifically, the \(x\) component of the spin polarization \(S_x\) still takes the form of Rabi oscillations, but unlike in the previous cases, its maximum value is almost the exact opposite of the initial value at \(t=0\), as shown in Fig.~\ref{fig9}(b1). At the resonance frequency \(\omega = \omega_1 = \omega_2\), resonance occurs between the states \(\ket{11}\) and \(\ket{21}\) as well as between \(\ket{12}\) and \(\ket{22}\), resulting in the state vector of the system at any time being given by:
	\begin{equation}\label{eq22}
		\begin{aligned}		
			\ket{\Psi(t)}=&\dfrac{1}{\sqrt{2}}(\cos(\lvert{ \mathcal{V}}\rvert t)e^{-iE_{12}t}\ket{12}-\cos(\lvert{ \mathcal{V}}\rvert t)e^{-iE_{11}t}\ket{11}\\&+i\sin(\lvert{ \mathcal{V}}\rvert t)e^{-iE_{21}t}\ket{21})+i\sin(\lvert{ \mathcal{V}}\rvert t)e^{-iE_{22}t}\ket{22}),
		\end{aligned}
	\end{equation}	
	where the coupling coefficient between \(\ket{11}\) and \(\ket{21}\) is \(\lvert{ \mathcal{V}_1}\rvert\), and between \(\ket{12}\) and \(\ket{22}\) is \(\lvert{ \mathcal{V}_2}\rvert\), with \(\lvert{ \mathcal{V}_1}\rvert \approx \lvert{ \mathcal{V}_2}\rvert = \lvert{ \mathcal{V}}\rvert\) [see Fig.~\ref{fig11}(a)]. The fact that the tuning of the system parameters can achieve \(\lvert{ \mathcal{V}_1}\rvert \approx \lvert{ \mathcal{V}_2}\rvert = \lvert{ \mathcal{V}}\rvert\) offers an explanation for why the two sets of resonances are perfectly coincident. According to Eq.~(\ref{eq22}), at \(t=\dfrac{(2n+1)\pi}{2\lvert{ \mathcal{V}}\rvert}\), the sign flipping of \(S_x\) occurs because the system has fully transitioned from its initial state, which is a superposition of \(\ket{11}\) and \(\ket{12}\), to a state formed by an equal-weight superposition of \(\ket{21}\) and \(\ket{22}\). Since \(\ket{21}\) and \(\ket{22}\) from the upper pair of levels have opposite \(x\)-directional spin polarizations compared to \(\ket{11}\) and \(\ket{12}\) from the lower pair of levels, we readily obtain \(S_{x}(t)=\dfrac{1}{2}\bra{\Psi(t)}\sigma_x\ket{\Psi(t)}\approx{-S_{x}(0)}\) at \(t=\dfrac{(2n+1)\pi}{2\lvert{ \mathcal{V}}\rvert}\) [the behavior can be seen in Fig.~\ref{fig9}(b1)]. Adhering to the same routine employed in the preceding two cases, we obtain the analytical $P_L(t)$ by substituting Eq.~(\ref{eq22}) into Eq.~(\ref{eq18}), as follows
	\begin{equation}\label{eq23}
		P_L(t)=\dfrac{1}{2}+\dfrac{1}{4}\{\cos[(2\lvert{ \mathcal{V}}\rvert+\Delta{E}) t]+\cos[(2\lvert{ \mathcal{V}}\rvert-\Delta{E})t]\},
	\end{equation}
		\begin{figure}[htp]
		\center
		\includegraphics[width=8cm]{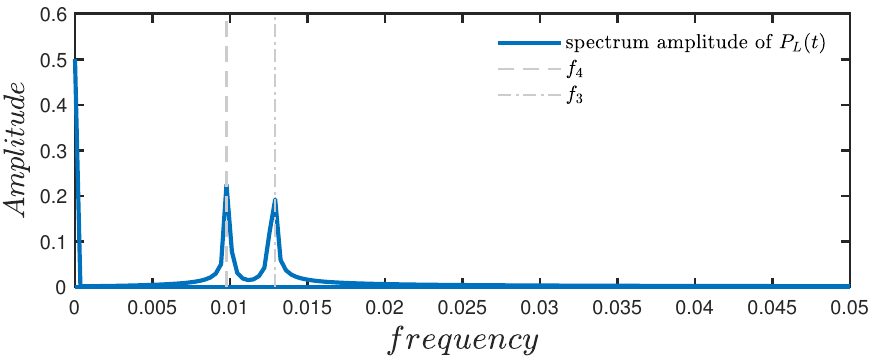}
		\caption{Frequency spectrum of $P_L(t)$ for the case in Fig.~\ref{fig9}(a1). The blue line depicts the result obtained from Fourier analysis of the numerical data for $P_L(t)$ in Fig.~\ref{fig9}(a1). The vertical (dashed and dash-dotted) lines in the figure correspond to two analytical frequencies obtained from Eq.~(\ref{eq23}).}
		\label{fig10}
	\end{figure}
	where $\Delta{E}=E_{12}-E_{11}=E_{22}-E_{21}$. From Eq.~(\ref{eq23}) it can be seen that \(P_L\) is a combination of two cosine oscillations with different frequencies \(f_3 = 2|\mathcal{V}| + \Delta{E}\) and \(f_4 = 2|\mathcal{V}| - \Delta{E}\). In Fig.~\ref{fig10}, we have conducted numerical analysis of the frequency spectrum of \(P_L(t)\) derived from Eq.~(\ref{eq4}), and we have observed that the spectrum displays two peaks, each situated precisely at the analytical frequencies \(f_3\) and \(f_4\). Due to the presence of the two different frequencies, the tunneling process is weakened, as evidenced by the modulation of the oscillatory amplitudes of \(P_L(t)\) in Fig.~\ref{fig9}(a1). Nevertheless, the long-time average of \(P_L(t)\) still approaches 0.5, as illustrated in Fig.~\ref{fig9}(a).
	
	\section{Tuning resonance transition by adjusting SOC}\label{IV}
	\subsection{SOC-modulated Resonance Transition Coupling Strength and Resonance Suppression}
\begin{figure}[htp]
		\center
		\includegraphics[width=8cm]{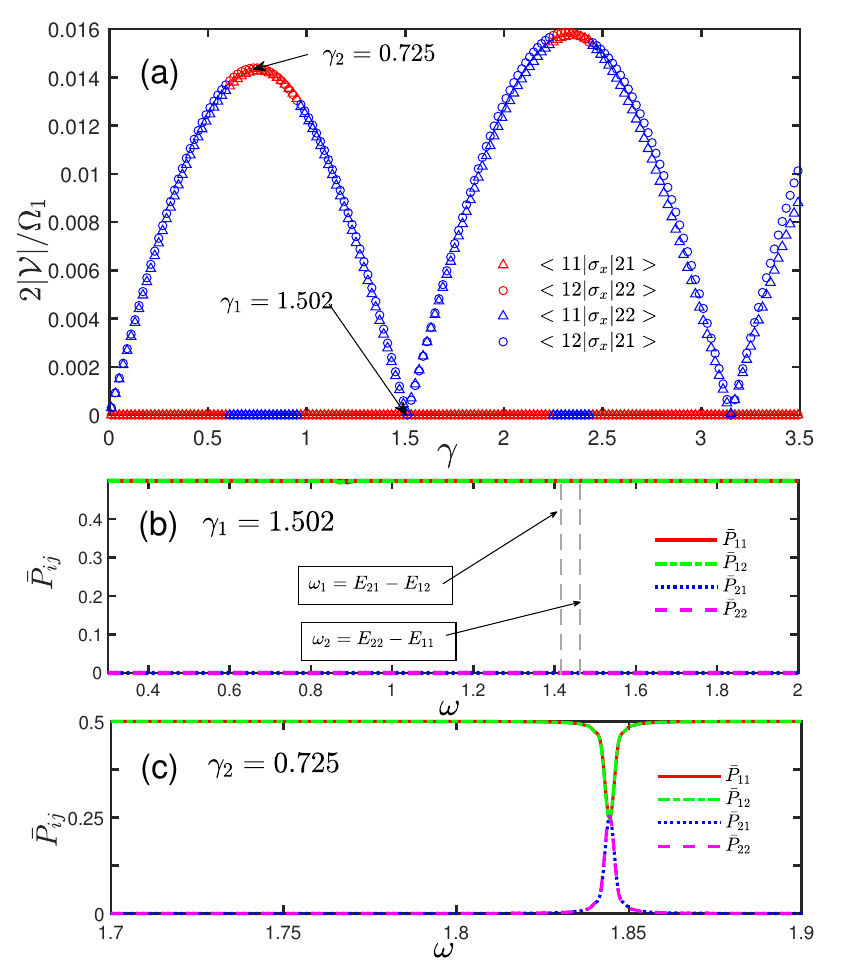}
		\caption{(a) The coupling coefficient $\lvert{\mathcal{V}}\rvert$ given by Eq.~(\ref{eq14}) is plotted against the SOC strength $\gamma$. (b) and (c) illustrate the dependence of $\bar{P}_{ij}$ on the driving frequency $\omega$, each corresponding to a specific value of $\gamma$ indicated by the arrows in (a). The parameter $\Omega_1 = 0.1$, and the other parameters are the same as those used in Fig.~\ref{fig2}. In (b), the dashed line indicates the frequency at which resonance was originally anticipated to occur. }
		\label{fig11}
	\end{figure}

	It is evident from Eq.~(\ref{eq14})  that the coupling coefficient $\lvert \mathcal{V} \rvert$ for the resonant transition is determined exclusively by the driving amplitude $\Omega_1$ and the matrix element $\bra{\alpha}\sigma_x\ket{\beta}$. After selecting the driving amplitude $\Omega_1$, the matrix element $\bra{\alpha}\sigma_x\ket{\beta}$—which relies exclusively on the unperturbed eigenstates—emerges as the unique factor governing the coupling coefficient $\lvert \mathcal{V} \rvert$ for the resonant transition. Given that SOC can profoundly influence the eigenenergies and $\mathcal{PT}$ symmetry of the unperturbed system, the correlation between the strength of SOC and the coupling coefficient $\lvert \mathcal{V} \rvert$ of the resonant transition is of great interest to us. In Fig.~\ref{fig11}(a), we numerically calculate the matrix element $\bra{\alpha}\sigma_x\ket{\beta}$ as a function of the SOC strength $\gamma$. As illustrated in Fig.~\ref{fig11}(a), the SOC strength demonstrates a controlling influence on the resonance transition coupling coefficient $\lvert \mathcal{V} \rvert$. It is established that the coupling coefficient of the resonant transition is inversely proportional to the duration of the transition between the two resonant states. That is, a larger resonance transition coupling coefficient $\lvert \mathcal{V} \rvert$ corresponds to a shorter transition period between the resonant states [cf. Eq.~(\ref{eq16})], indicative of a stronger resonance, and vice versa.
	Consequently, the resonance transition period can be finely tuned by modulating the SOC strength $\gamma$. Specifically, as depicted in Fig.~\ref{fig11}(a), there are particular values of $\gamma$ at which the coupling coefficient $\lvert \mathcal{V} \rvert$ vanishes, indicating that the resonance transition between energy levels induced by external driving is entirely eliminated. This demonstrates the capability of SOC to effectively quench resonance transitions between quantum states that are initiated by external perturbations. In Fig.~\ref{fig11}(a), the positions linked by different colored (red and blue) markers correspond to the points of abrupt changes in the $\mathcal{PT}$ symmetry in Fig.~\ref{fig2}(b). In Figs.~\ref{fig11}(b) and (c), we show the time dependence of $P_{ij}$ on the driving frequency $\omega$ at the two special SOC strength $\gamma_1 = 1.502$ and $\gamma_2 = 0.725$ as indicated in Fig.~\ref{fig11}(a), respectively. We note that due to $\gamma_1 = 1.502$ leading to $\lvert \mathcal{V} \rvert = 0$, the expected resonance peak at the gray dashed lines in Fig.~\ref{fig11}(b) is absent. Conversely, with $\gamma_2 = 0.725$ resulting in $\lvert \mathcal{V} \rvert \neq 0$, the resonance transition occurs at a certain modulation frequency, indicated by the peak in Fig.~\ref{fig11}(c).
	
	\subsection{ SOC-mediated Transition from Multiphoton to Fundamental Resonance }
\begin{figure}[htp]	
		\center
		\includegraphics[width=8cm]{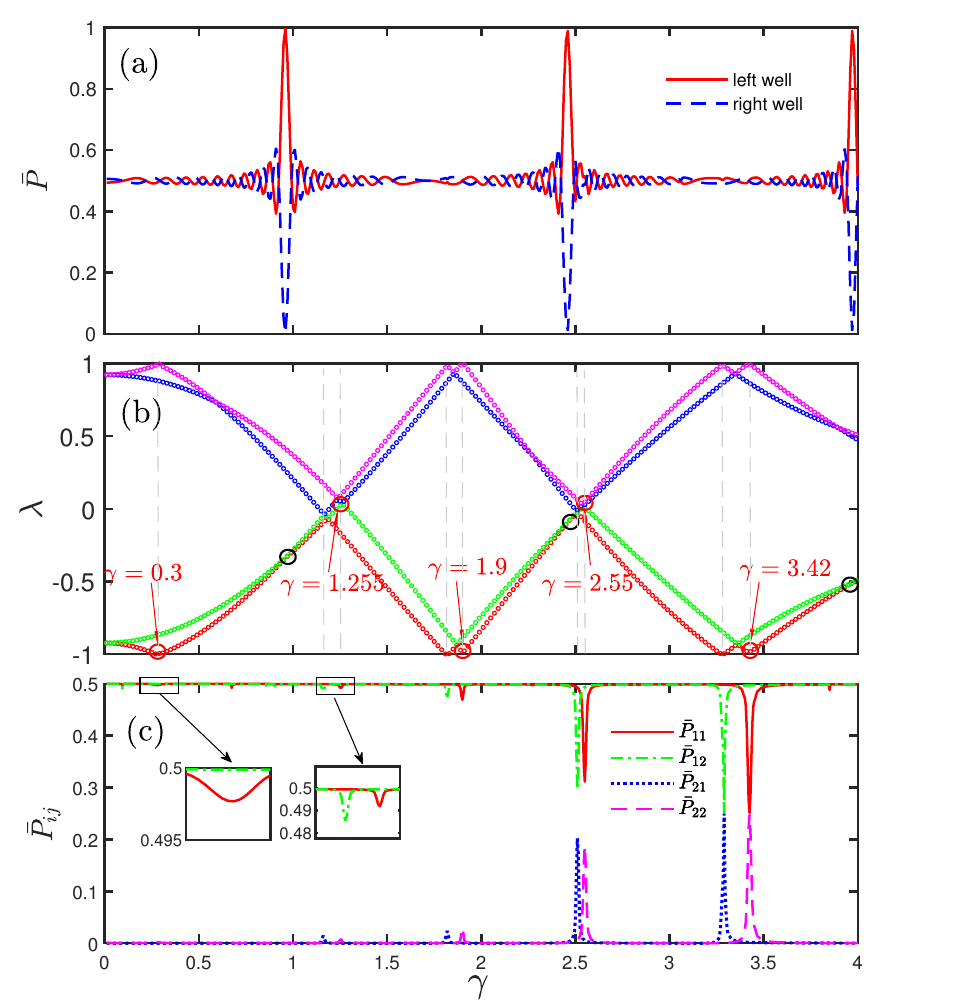}
		\caption{(a) The time-averaged probability $\bar{P}$ of finding the atom in the left (or right) well, (b) the quasienergies $\lambda$, and (c) $\bar{P}_{ij}$ are plotted against the SOC strength $\gamma$ at the fixed modulation frequency $\omega = 0.4$. The other parameters are the same as those in Fig.~\ref{fig11}. For better visualization, the quasienergies are rescaled by a factor of $\omega/2$, placing them within the first Brillouin zone with a range of (-1, 1]. The quasienergies $\lambda$ are numerically obtained from the simplified four-state model (\ref{eq8}), and the results in (a) and (c) are from the numerical simulation of the original Schr\"{o}dinger equation (\ref{eq4}). The inset in (c) provides an enlarged view of the resonance transition of $\bar{P}_{ij}$ at $\gamma =0.3$ and $\gamma = 1.255$.}
		\label{fig12}
	\end{figure}

	In this subsection, we delve further into examining the effects of SOC on the dynamics of the driven system. In Fig.~\ref{fig12}(a), we present the calculated time-averaged probability of finding the atom in the left and right wells as a function of $\gamma$, with a fixed modulation frequency of $\omega=0.4$. The system parameters are consistent with those used in Fig.~\ref{fig2}, and the initial state remains prepared as $\ket{1-}$. Fig.~\ref{fig12}(a) reveals a series of localization peaks (with  $\bar{P}_L=1$) at specific values of $\gamma\approx 0.96, 2.46, 3.97$, indicating that the atom is always localized in the left well. To deepen our understanding of these dynamical phenomena, we have depicted the quasienergies, labeled as $\lambda$, for the Floquet system in Fig.~\ref{fig12}(b). These quasienergies are calculated numerically by diagonalizing the evolution operator over one driving period, based on the four-state model (\ref{eq8}). Due to the quasienergy spectrum of a periodically time-dependent quantum system possessing a Brillouin zone-like structure, with the width of one zone being $\omega$, we therefore restrict the quasienergy to the first Brillouin zone, $\lambda/(\omega/2) \in (-1,1]$.
	It is clearly seen that the peaks of localization correspond exactly to the quasienergy crossings in the lower pairs (denoted by black open circles), matching the locations where $E_{12}-E_{11}=0$ for the unperturbed system, as observed in Fig.~\ref{fig2}(a). Thus, we can induce degeneracy at the quasienergy level and thus localize the atom by manipulating the strength of the SOC. Additionally, from the quasienergy spectrum, we can detect a sequence of avoided crossings between the upper and lower pair states, which are indicative  of photon-like resonances. To better  visualize these resonances, in Fig.~\ref{fig12}(c) we display the dependence of the time-averaged probability ${\bar{P}_{ij}}$ of finding the system in each unperturbed eigenstate $\ket{ij}$ on the SOC strength $\gamma$. We find that the peaks (and dips) in this quantity are indeed centered on the points where avoided crossings of quasienergies  take place. Furthermore, the resonance becomes more and more stronger, as manifested by the increasing height (and depth) of the corresponding resonance peaks (and dips) in ${\bar{P}_{ij}}$ as the SOC strength $\gamma$ is enhanced.

	In Fig.~\ref{fig12}(b),the avoided crossings between the upper and lower pair states, which corresponds to the turning points of the quasienergies near the center $\lambda = 0$ and at the boundary of the Brillouin zone,
	\begin{figure}[htp]	
		\center
		\includegraphics[width=8.8cm]{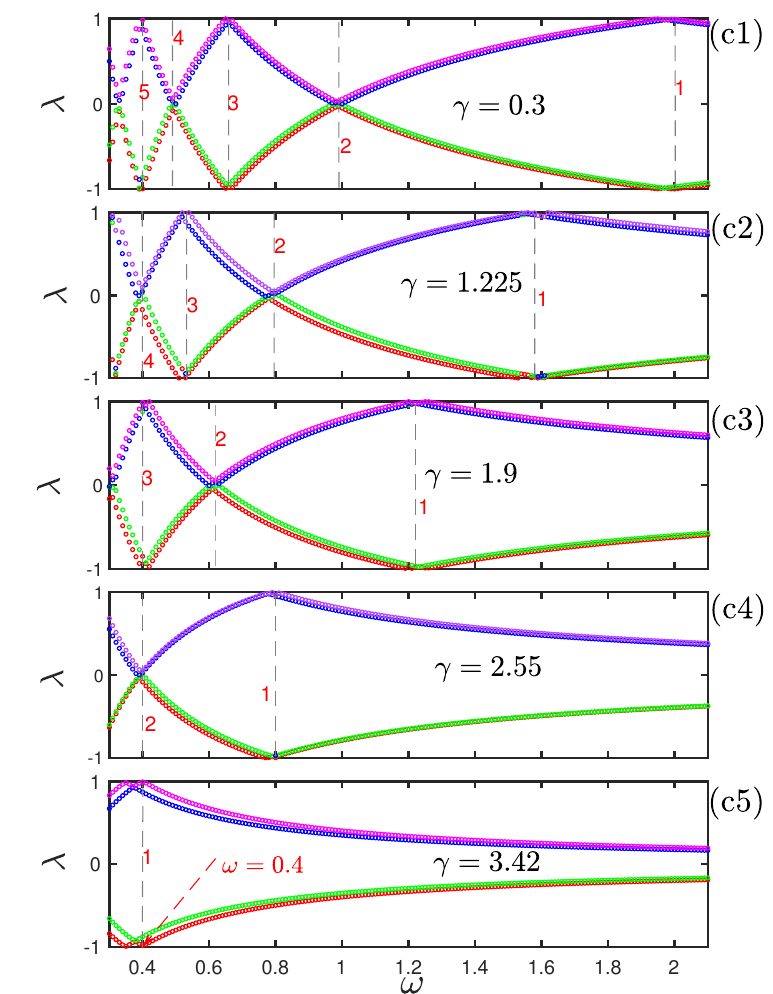}
		\caption{ Dependence of quasienergies $\lambda$ on the modulation frequency $\omega$ at the different values of SOC strength $\gamma$, marked by the red circles in Fig.~\ref{fig12}(b). Each resonance (vertical line) is labeled with the number of photons involved in the process. The parameters are the same as those used in Fig.~\ref{fig11}. }
		\label{fig13}
	\end{figure}
	are marked by the gray dashed lines. These gray dashed lines indicate the points where resonance transitions occur between the states $\ket{11}$ and $\ket{22}$, or between the states $\ket{12}$ and $\ket{21}$, as manifested by the corresponding transition of $\bar{P}_{ij}$ shown in Fig.~\ref{fig12}(c). To better examine these resonances, in Fig.~\ref{fig13}, we illustrate the dependence of the quasienergies on the modulation frequency $\omega$ for the specific values of SOC strength $\gamma$, marked by the red open circles in Fig.~\ref{fig12}(b). Fig.~\ref{fig13} reveals a sequence of $m$-integer photon-assisted resonances at various modulation frequencies for each fixed SOC strength, with the numbers in the graphs denoting the order of the associated multiphoton resonances. We have confirmed that the condition $m\omega = E_{22} - E_{11}$ reproduces well the positions of the one-photon, two-photon, three-photon, four-photon, and five-photon-like resonances. For the SOC strength $\gamma=0.3, 1.225, 1.9, 2.55, 3.42$, the location $\omega= 0.4$ corresponds to the 5, 4, 3, 2, and 1 photon peaks, respectively.  This indicates that the increase of SOC leads to a transition from multiphoton resonance (higher-order effect) to fundamental resonance (first-order effect), thereby enhancing the resonance strength. The fundamental reason for this phenomenon lies in the fact that increasing the SOC strength can correspondingly reduce the energy gap between the lower and upper pairs of energy levels, thus requiring a smaller number of  photons to bridge this energy difference. Therefore, the adjustment effect of SOC on the energy levels provides us with a richer means to manipulate quantum dynamics and enhance resonances.
	\section{Conclusion}\label{V}
	In this study, we have primarily investigated the effect of SOC on the resonance transition triggered by periodically modulated Raman coupling for a single boson confined in a symmetric double-well potential. We have used Floquet theory to derive analytical results for the resonance transition, providing a transparent approach to manipulate the tunneling and spin dynamics through the PAT mechanism. Considering the significant influence of SOC on the energy levels and $\mathcal{PT}$ symmetry of the unperturbed system, we can achieve resonance transitions between the pre-prescribed energy levels. This can trigger a transition from localization to full Rabi oscillation between the two potential wells, or effectively weaken the tunneling effect, manifesting as a quantum beating phenomenon. Additionally, such resonance transitions have the potential to induce spin flipping in a spin-orbit coupled atom. Furthermore, we have discovered that adjusting the SOC strength allows for precise control over the coupling coefficients of the resonance transition, with the capability to completely suppress the resonance. We have also observed that tuning the SOC can lead to a transition from multiphoton to fundamental resonance, markedly intensifying the resonance transition. This offers additional techniques and tools to mitigate the complex dynamics induced by resonance in experimental settings, or to enhance the resonance effect as necessary.  Our numerical results for the continuous model are all reproducible by analyzing a simplified four-state model, which allows us to gain a deeper understanding of the underlying mechanisms and dynamics of the system and to make accurate predictions.
	
	\acknowledgments
	The work was supported by the National Natural Science Foundation of China (Grants No. 12375022 and No. 11975110), the Natural Science Foundation of Zhejiang Province (Grant No. LY21A050002), and Zhejiang Sci-Tech University Scientifc Research Start-up Fund (Grant No. 20062318-Y).

\end{document}